\documentstyle[12pt]{article}
\begin{document}

\begin{flushright}
MPI-PhT/96-76\\
August 1996\\
\end{flushright}
\begin{center}
\large {\bf A Short Review and Some New Results on Polarization}
\footnote{ Talk given at the 1996 Montpellier Conference on QCD}
\\
\mbox{ }\\
\normalsize
\vskip3cm
{\bf Bodo Lampe}                          
\vskip0.3cm                               
Max Planck Institut f\"ur Physik \\       
F\"ohringer Ring 6, D-80805 M\"unchen \\  
\vspace{3cm}                              

{\bf Abstract}\\                          
\end{center}                              

At the moment I am writing a long review \cite{lampereya}
on polarization in deep inelastic scattering. Therefore 
I thought it might be a good idea to present the highlights
here. Some of my own results will be discussed in the end.

\newpage

One of the main issues in polarized DIS experiments is the 
question of how the proton spin at high energies is composed 
out of the spins of its constitutents, possibly 
\begin{equation}
+{1\over 2} = {1\over 2} \Delta\Sigma + \Delta g + L_z \label{111}
\end{equation}
where $\Delta\Sigma = \Delta (u+ \bar{u}) + \Delta (d+ \bar{d}) 
+ \Delta (s+ \bar{s})$ is the contribution from the quark spins. 
In the (static) constituent models, like SU(6), one has 
$\Delta\Sigma (SU_6)=1$ but experimentally it seems that this 
rule is violated by a large amount ($\Delta\Sigma_{exp} \approx 0.25$). 
This experimental fact is also in disagreement with the rather crude 
Ellis-Jaffe sum rule \cite{ellisjaffe} which predicts 
$\Delta\Sigma_{EJ} \approx 0.65$ on the basis of the approximation 
$\Delta s = \Delta \bar{q} =\Delta g=0$. It will probably turn out 
that both the gluon and the strange quark are needed to understand 
the proton spin structure. 

In general, one is not only interested in the first moment 
$\Delta g$, but also in the x--dependence of the polarized gluon 
density $\delta g(x)$, although I think that $\Delta g$ has a 
particular significance, because it gives the gluon contribution 
to the spin ${1\over 2}$ of the proton and, as we shall see, is 
dynamically related to the anomaly \cite{lampealtarelli}. 
The present experiments do not really give good information 
about $\delta g(x)$, because in DIS the gluon is a higher order effect. 
I shall come back to the question of determining $\delta g(x)$ at the 
end of this talk. 

In principle, to determine the first moment 
$\Delta g=\int_0^1 dx \delta g(x)$ very precisely it is necessary to 
know the small--x behaviour. From the theoretical side, it can be 
said that the polarized structure function behaves much more 
moderate than the unpolarized one. Although this question is not 
quantitatively settled \cite{ryskin}, \cite{vogt} one can get qualitative 
insight by inspection of the AP evolution kernels. Whereas the 
unpolarized kernel $P_{gg}$ has a ${1 \over x}$ singularity for 
$x \rightarrow 0$ (corresponding to a ${1 \over x}$ singularity 
of the unpolarized structure function $F_2$), the polarized kernels 
are all perfectly finite suggesting a more or less constant behaviour 
of $g_1$ for $x \rightarrow 0$. 
Some authors \cite{ryskin} have argued in favor of a somewhat more 
singular behavior, but the singularities are certainly integrable 
and would not modify the value of $\Delta g$ too much.  
I do not want to discuss these approaches in further detail here 
because they are incomplete and not fully consistent. 

The most important and precise 
information about the proton spin comes from the CERN 
data \cite{adams}. These measurements essentially have been 
confirmed at SLAC. There are also neutron and deuteron data available. 
They can be used to check the Bjorken sum rule which is a nonsinglet 
sum rule. However, these data have larger errors and do not give 
additional information about the singlet quantities 
$\Delta\Sigma$ and $\Delta g$. 

The total of present measurements have been used to extract informations 
about the parton densities. In particular the valence quark densities 
$\delta u_v(x)$ and $\delta d_v(x)$ are now known to some accuracy 
\cite{gehrmann}, \cite{vogelsang}. In contrast to the valence quarks, 
the polarized sea quarks and gluons are hardly fixed, because they enter 
the analysis only as a higher order effect. It has been attempted to 
determine the magnitude of the first moment $\Delta g$ \cite{forte}, 
which has a better chance because it is an average. As an essential 
input the presently known $Q^2$--dependence of the data (both theoretically 
and experimentally) was used, in which the polarized gluon plays some 
role. On the basis of this, a value $\Delta g \approx 1.3 \pm 0.5$ was 
quoted. My opinion is that the error here is underestimated but the 
order of magnitude of $\Delta g$ looks reasonable. 
To determine the polarized gluon density, other experiments   
will probably be superior to DIS, like eg. direct photon production 
at RHIC ($\delta q \otimes \delta g \rightarrow \gamma^{\ast}q$) 
or heavy quark production. I shall come back to this issue later. 

In all the analyses \cite{gehrmann}, \cite{vogelsang} and \cite{forte} 
the AP splitting functions have been used which are now known to 
2--loop accuracy \cite{mertig}. As is well known, 2--loop 
splitting functions are needed for a 1--loop(=NLO) analysis of 
structure functions. Although the corrections $\delta P_{ij}^{2loop}$ 
to the leading order $\delta P_{ij}^{1loop}$ appearing in the 
Altarelli--Parisi equation  
\begin{equation}
\frac{d}{dt} {\Delta\Sigma (x,Q^2)\choose \Delta g(x,Q^2)}=
(\delta P^{1loop}+\delta P^{2loop}) 
\otimes {\Delta\Sigma\choose \Delta g}
\label{4213}
\end{equation}
turn out to be quite 
small numerically, they are important to know from a principle 
point of view because they essentially fix the ambiguities from 
higher orders. 

The calculations of the $\delta P_{ij}^{2loop}$ have been 
done in the $\overline{MS}$--scheme. This has the advantage that 
the calculation is manageable and furthermore in alignment with 
with the unpolarized analysis which is also usually done in the 
$\overline{MS}$--scheme. Unfortunately, the treatment of 
$\gamma_5$ (not present in the unpolarized case) in these 
calculations is such that the contribution from the anomaly is not 
transparent. \footnote{It would be desirable to have a prescription 
of $\gamma_5$ in dimensional regularization in which the contribution 
from the anomaly becomes explicit. I do not know whether such a 
prescription exists, but it would certainly be valuable.\label{fn1}} 
Let me explain how this comes about: In the framework of the 
operator expansion, the connection between polarized DIS and 
the anomaly can be easily established. Namely, the matrix element  
of the anomalous axial vector singlet operator 
$j_{\mu ,S}^5=\sum_q \bar{q} \gamma_{\mu} \gamma_5 q$ appears as 
the leading twist contribution to the first moment of the singlet 
component of $g_1$, i.e. 
\begin{equation}
\int_0^1 dx g_{1,S}^p(x,Q^2)
\sim <PS\vert j_{\mu ,S}^5 \vert PS>
\label{42}
\end{equation}
This well known fact has been used in many analyses in the past, eg. 
recently by \cite{fritzsch}. 
Note that the singlet part $\int_0^1 dx g_{1,S}^p$ can be 
obtained from the experimentally available $\int_0^1 dx g_{1}^p$, 
because the nonsinglet parts are known due to low energy information 
from $\beta$--decays. 

Due to its anomaly, the matrix element of $j_{\mu ,S}^5$ has 
an anomalous component, 
\begin{equation}
 <PS\vert j_{\mu ,S}^5 \vert PS> \sim \Delta\Sigma - n_f 
\frac{\alpha_s}{2\pi}\Delta g(Q^2)
\label{43}
\end{equation}
where the first term $\Delta\Sigma$ is the ordinary quark--photon 
scattering contribution and the second term $\sim \Delta g$ is 
the contribtuion from the anomaly. Unfortunately, the representation 
(\ref{43}) depends on the regularization scheme. It is true only in 
suitable chirality preserving schemes where transitions between quarks 
of different helicities are forbidden to any order in perturbation 
theory. In such schemes $\Delta\Sigma$ is $Q^2$--independent. 
In the $\overline{MS}$--scheme used by \cite{mertig} with their 
particular $\gamma_5$ prescription, chirality is not conserved and 
one has $ <PS\vert j_{\mu ,S}^5 \vert PS> \sim \Delta\Sigma(Q^2)$, 
i.e. the anomalous piece has effectively been absorbed into a 
redefintion of $\Delta\Sigma$ (see the footnote). 

In the previous paragraphs the reader should have realized, among other 
things, the 
difficulties to extract the polarized gluon distribution 
and in particular its first moment 
from inclusive deep inelastic data. 
These problems have been anticipated several years ago 
by theoretical studies \cite{altarellistirling,carlitzreyaaltarelli}, 
and they are in fact not surprising in view of the 
subtleties in determining the unpolarized gluon density 
in unpolarized DIS experiments \cite{herastudies}. 

A popular way out of this dilemma is the study of 
semi--inclusive cross sections, and in particular of 
charm production, because the production of heavy quark hadrons  
is triggered in leading order by the photon--gluon fusion 
mechanism and is therefore sensitive to the gluon 
density inside the proton, whereas the heavy quark 
content of the proton is usually negligible at presently 
available $Q^2$--values. 

Due to its prominent decay mechanism, $J/\psi$--events 
are the most prominent among the charm quark production, 
and this fact has led to attempts to determine the 
unpolarized gluon density from the $J/\psi$--production 
cross section \cite{bergerjones,martinryskin}. 
Similar ideas hold in the case of polarized $J/\psi$--production. 
However, these suggestions are model dependent and 
depend on assumptions which go beyond the QCD improved 
parton model. For example, according to the suggestion 
of \cite{martinryskin} elastic $J/\psi$--production 
should measure the square of $\delta g(x)$   
and therefore be 
very sensitive to its magnitude.  
However, it is not clear whether in 
the cross section formula there is a factor $g(x_1)g(x_2)$ 
or whether some independent 2--gluon correlation function $K(x_1,x_2)$ 
appears.    
Furthermore, it has recently been stressed \cite{braaten}, 
that color octet contributions may appear in addition 
to the color singlet pieces in inelastic $J/\psi$--production. 
If true, this would upset the inelastic $J/\psi$--analysis 
because several new free 
parameters, the color octet matrix elements, would enter the game. 

Therefore, from the theoretical point of view 
the cleanest signal for the gluon in heavy quark 
production is probably open charm production, although 
experimentally it has worse statistics due to the  
difficulties in identifying D--mesons. 
Instead of the deep inelastic process one may as well look 
at photoproduction, because the
mass of the charm quark forces the process to take 
place in the perturbative regime. The advantage 
of photoproduction over DIS is its larger cross section. 
Two fixed target experiments, COMPASS at CERN and 
an experiment at SLAC are 
being developed to measure the polarized gluon 
distribution via photoproduction.  

In leading order the inclusive 
polarized deep 
inelastic open charm production cross section is given by 
\cite{reyavogelsang,watson} 
\begin{equation}
{d \Delta \sigma_c \over dQ^2 dy}={4\pi \alpha^2 \over Q^2} 
{2-y \over yS} g_1^c({Q^2 \over yS},Q^2) 
\label{650}
\end{equation}
where 
\begin{equation}
g_1^c(x,Q^2)={\alpha_s \over 9\pi}\int_{(1+{4m_c^2 \over Q^2})x}^1 
{dw \over w} \delta g(w,Q^2) \delta h({x \over w}) 
\label{651}
\end{equation}
is the charm contribution to the polarized structure 
function $g_1$ and where  
\begin{equation}
\delta h(z)=(2z-1)\ln {1+\beta \over 1-\beta}+(3-4z)\beta 
\label{652}
\end{equation}
is the parton level matrix element. One has  
$\beta=\sqrt{1-{4m_c^2 \over \hat s}}$ where the Mandelstam 
variable $\hat s$ is defined by $\hat s=(p+q)^2=Q^2{1-z \over z}$. 
By combining these formulae with the unpolarized cross 
section one can obtain 
the polarization asymmetries $d \Delta \sigma^c \over d \sigma^c$. 
If one plugs in the relatively large gluon contribution 
of reference \cite{altarellistirling}, one gets asymmetries 
of the order 0.1 in a fixed target experiment which would 
operate well above charm threshold. 

It is straightforward to obtain from the above expressions 
(\ref{650}) -- (\ref{652}) 
the inclusive open charm photoproduction cross section 
by taking the simultaneous limits $Q^2 \rightarrow 0$ and 
$z \rightarrow 0$ while keeping ${Q^2 \over z} \approx \hat s$ 
fixed: 
\begin{equation}   
\Delta \sigma^c_{\gamma p} (S_{\gamma})=
{8 \pi \alpha \alpha_s \over 9 S_{\gamma}}
\int_{{4m_c^2 \over S_{\gamma}}}^1 
{dw \over w} \delta g(w, S_{\gamma})   
(3v-\ln {1+v \over 1-v})   
\label{653}
\end{equation}     
where $v=\sqrt{1-{4m_c^2 \over \hat s}}$ and $\hat s=wS_{\gamma}$. 
This integrated cross section depends only on the 
total proton--photon energy $S_{\gamma}=(P+q)^2$ which for 
a fixed target experiment is given by $S_{\gamma}=2ME_{\gamma}$ 
where $E_{\gamma}$ is the photon energy. By varying the 
photon energy it is in principle possible to explore the 
x--dependence of $\delta g$. Very high photon energies 
correspond to small values of x. However, as we shall see 
later, it is not trivial to obtain the first moment 
of $\delta g$ from the cross section Eq. (\ref{653}). 

It should be noted that the second argument of 
$\delta g$ in Eqs. (\ref{651}) and (\ref{653}) is not 
certain. It might as well be $4m_c^2$ or any number in 
between. This uncertainty 
reflects our ignorance about the magnitude of the 
higher order correction and could be resolved if a 
higher order calculation of these cross sections would be 
performed.  
The same statement holds true for the argument 
of $\alpha_s$. Therefore, in the equations presented below 
the energy arguments of $\delta g$ and $\alpha_s$ will be chosen 
to be more general, $\mu_S$ and $\mu_R$ respectively.  

Eq. (\ref{653}) was obtained after integration over the 
charm quark production angle ($\hat \theta$ in the gluon photon cms). 
If one is interested in the 
$p_T$ distribution or wants to introduce a $p_T$--cut, it is 
appropriate to keep the $\hat \theta$ dependence in the 
fully differential cross section 
\begin{equation}  
d \Delta \sigma^c_{\gamma p} = {e_c^2 \alpha_s(\mu_R) \over 16\hat s} 
 dw \delta g(w,\mu_S) 
vd \cos  \hat \theta     
(2 {\hat t^2+ \hat u^2 -2m_c^2\hat s \over \hat t\hat u}
+4m_c^2{\hat t^3+\hat u^3 \over \hat t^2\hat u^2})   
\label{654}
\end{equation}
where $\hat s=wS_{\gamma}$, 
$\hat t=-{\hat s\over 2}(1-v\cos \hat \theta)$ and 
$\hat u=-{\hat s\over 2}(1+v\cos \hat \theta)$. 
It is possible to make a transformation to the transverse 
charm quark momentum by using 
$p_T^2=({\hat s \over 4}-m_c^2)\sin^2\hat \theta$: 
\begin{equation}  
\sigma(p_{Tcut})={e_c^2 \alpha_s(\mu_R) \over 16 S_{\gamma}} 
\int_{{4m_c^2 \over S_{\gamma}}}^1 {dw \over w} 
\delta g(w,\mu_S) {v \over {\hat s \over 4}-m_c^2}    
\int_{p_{Tcut}^2}^{{\hat s \over 4}-m_c^2} 
{d p_T^2 \over \sqrt{1-{p_T^2 \over {\hat s \over 4}-m_c^2}}} 
(2m_c^2{\hat s \over \hat t\hat u}-{\hat t \over \hat u} 
                                    -{\hat u \over \hat t}  
-2m_c^2({\hat t \over \hat u^2}+{\hat u \over \hat t^2}))  
\label{655}
\end{equation}  
There are several good reasons to study the $p_T$ distribution. 
First of all and in general, it gives more information than the 
inclusive cross section. Secondly and in particular, it can be 
shown that the integrated photoproduction cross section 
Eq. (\ref{653}) as well 
as the corresponding 
DIS charm production cross section are not sensitive 
to the first moment of $\delta g$. The sensitivity is strongly 
increased, however, if a $p_T$--cut of the order of 
$p_T \geq 1$ GeV is introduced. 
Last but not least, 
it is experimentally reasonable to introduce a $p_T$--cut. 

Let us dwell on the first moment discussion for a moment. 
It is true that the first moment is only one among an infinite 
set of moments and the most interesting quantity to know is the full 
x--dependence of $\delta g$. However, 
the first moment $\Delta g$ certainly has its 
significance, because it enters the fundamental spin 
sum rule (\ref{111}) and because it gives the contribution 
within the proton to the $\gamma_5$ anomaly. 

In massless DIS it is straightforward to find out what the 
contribution of the first moment to the cross section is. 
One can apply the convolution theorem to see that the 
contribution of $\Delta g$ is given by the first moment of the 
parton matrixelement. 
If masses are involved, like $m_c$, the  
answer to this question is somewhat more subtle. Since the cross 
section is not any more a convolution of the standard form 
one has to write it artificially as 
$\sigma (a,e)=\int_a^1 {dw \over w} \delta g(w) H({a\over w},e)$ 
where $a={4m_c^2 \over S_{\gamma}}$ for photoproduction and 
$a=(1+{4m_c^2 \over q^2})x$ for DIS charm production 
and e stands for external variables like  
$e=p_{Tcut}^2, S_{\gamma}, ...$ etc. Now one can apply the 
convolution theorem and the first moment 
$\int_0^1 dzH(z,e)$ 
gives essentially the contribution from $\Delta g$. 
It turns out that both for the inclusive charm photoproduction 
and DIS (Eqs. (\ref{653}) and (\ref{650})) the corresponding 
quantities $\int_0^1 dzH(z,e)$ identically vanish \cite{lampegambino}. 
This can be traced back to the
small--$p_T$ behaviour of the marixelement for 
$\gamma g \rightarrow c \bar c $
which cancels the contribution of the large--$p_T$ region
in $\int_0^1 dzH(z,e)$ \cite{schaefermankiewics}.
It is 
not really a surprise in view of the structure of the anomaly 
in massive QCD (cf. the appendix of ref. \cite{lampe4}). 

Since the integrals $\int_0^1 dzz^{n-1}H(z,e)$ keep being small 
in a neighbourhood of n=1 one may conclude from this that 
these cross sections are not suited for determining the first 
moment of $\delta g$. Fortunately, the situation changes 
drastically if one includes a $p_T$--cut of greater than 1 GeV. 
In that case the sensitivity 
to $\Delta g$ is re--established because the  
small--$p_T$ behaviour of the marixelement for 
$\gamma g \rightarrow c \bar c $ 
does not cancel the contribution of the large--$p_T$ region 
any more. In ref. \cite{lampegambino} a more quantitative 
study of these facts will be carried out and the sensitivity 
of the above mentioned RHIC process to $\Delta g$ 
will be analysed as well.

\end{document}